\documentclass[conference]{IEEEtran}
\IEEEoverridecommandlockouts
\usepackage{cite}

\usepackage{balance}

\usepackage{amsmath,amssymb,amsfonts}
\usepackage{algorithmic}
\usepackage{graphicx}
\usepackage{subfigure}
\usepackage{multirow}
\usepackage{textcomp}
\usepackage{xcolor}
\usepackage{url}

\usepackage{enumitem}

\usepackage{tcolorbox}

\usepackage[framemethod=tikz]{mdframed}

\mdfdefinestyle{mpdframe}{
    frametitlebackgroundcolor   =black!15,
    frametitlerule    =true,
    roundcorner       =2pt,
    middlelinewidth   =1pt,
    innermargin       =0.1cm,
    outermargin       =0.1cm,
    innerleftmargin   =0.1cm,
    innerrightmargin  =0.1cm,
    innertopmargin    =0.1cm,
    innerbottommargin =0.1cm
}

\def\BibTeX{{\rm B\kern-.05em{\sc i\kern-.025em b}\kern-.08em
    T\kern-.1667em\lower.7ex\hbox{E}\kern-.125emX}}
\begin{document}

\title{A Search-Based Testing Framework for Deep Neural Networks of Source Code Embedding
}

\author{
    \IEEEauthorblockN{Maryam Vahdat Pour\IEEEauthorrefmark{1}, Zhuo Li\IEEEauthorrefmark{2}, Lei Ma\IEEEauthorrefmark{2}, Hadi Hemmati\IEEEauthorrefmark{1}}
    \IEEEauthorblockA{\IEEEauthorrefmark{1}Schulich School of Engineering, University of Calgary, Canada
    }
    \IEEEauthorblockA{\IEEEauthorrefmark{2}Department of Information Science and Electrical Engineering, Kyushu University, Japan
    }
    \IEEEauthorblockA{
    \{maryam.vahdatpour, hadi.hemmati\}@ucalgary.ca\\
    li.zhuo.786@s.kyushu-u.ac.jp, malei@ait.kyushu-u.ac.jp}
    
}


\maketitle
\thispagestyle{plain}
\pagestyle{plain}

\begin{abstract}

Over the past few years, deep neural networks (DNNs) have been continuously expanding their real-world applications for source code processing tasks across the software engineering domain, {\it e.g.}, clone detection, code search, comment generation. Although quite a few recent works have been performed on testing of DNNs in the context of image and speech processing, limited progress has been achieved so far on DNN testing in the context of source code processing, that exhibits rather unique characteristics and challenges.

In this paper, we propose a search-based testing framework for DNNs of source code embedding and its downstream processing tasks like Code Search. To generate new test inputs, we adopt popular source code refactoring tools to generate the semantically equivalent variants. For more effective testing, we leverage the DNN mutation testing to guide the testing direction. To demonstrate the usefulness of our technique, we perform a large-scale evaluation on popular DNNs of source code processing based on multiple state-of-the-art code embedding methods ({\it i.e.}, Code2vec, Code2seq  and CodeBERT). The testing results show that our generated adversarial samples can on average reduce the performance of these DNNs from 5.41\% to 9.58\%. Through retraining the DNNs with our generated adversarial samples, the robustness of DNN can improve by 23.05\% on average. The evaluation results also show that our adversarial test generation strategy has the least negative impact (median of 3.56\%), on the performance of the DNNs for regular test data, compared to the other methods.
\end{abstract}


\section{Introduction}

Recently, deep neural networks (DNNs) have been successfully applied to many application domains such as medical imaging, autonomous driving, and natural language processing (NLP). At the same time, there is also an increasing trend of adopting DNN to diverse source code processing tasks in the software engineering domain, {\it e.g.}, ``code search'', ``comment generation'', and ``program repair''. 
An essential stage of these tasks by DNN is to represent (encode) source code snippets into a vector representation called ``Code Embedding'' \cite{alon2019Code2vec}.
Ideally, two vectorized source code snippets with similar functionalities, are encoded into two close vectors so that certain code semantics are extracted and preserved during embedding.

Given the growing number of downstream tasks that rely on code embedding models, the performance of these models in terms of accuracy and robustness is critical. Though accuracy has always been the main objective, the robustness of code embedding models has not been studied much in the literature but highly desired.
A non-robust DNN may infer a completely different result even the input source code slightly changes.
For example, Ramakrishnan {\it et al.}, \cite{ramakrishnan2020semantic} show that the behavior of a code captioning model, which leverages a state-of-the-art code embedding model (Code2seq \cite{alon2018Code2seq}), changes its prediction after the simple insertion of a logging print statement, which does not change the code semantics.
In addition, the robustness issues can also pose potential security risks for the downstream task based on a DNN malware classifier, which can be abused by an attacker\cite{alon2018Code2seq}. 

In this paper, we focus on the adversarial robustness of code embedding models by (a) proposing a new search-based testing framework for generating adversarial examples for DNNs of source code processing, and (b) improving their robustness by retraining with our augmented adversarial tests. In general, the adversarial robustness of DNN models has been extensively studied in the literature, which generates test data (called adversarial samples) with small input perturbations to fool a DNN. 
Although adversarial attack methods for natural language process have been proposed \cite{goodfellow2014explaining, papernot2016limitations}, they are not directly applicable to source code, since the source code must strictly follow the language grammar, otherwise, the generated test is uncompilable and invalid.

In the literature, there are a few studies that propose specialized adversarial generation strategies for code embedding. For instance, both Rabin {\it et al.}, \cite{rabin2020evaluation} and Ramakrishnan {\it et al.}, \cite{ramakrishnan2020semantic} studies suggest using program refactoring operators to generate adversarial examples for source codes. Although our proposed adversarial code example generator is also based on refactoring operators, we propose to adopt mutation testing-based guidance for a more effective generation.
Furthermore, unlike the previous work, we improve the robustness of the models by retraining them using the adversarial examples and show how much improvement this provides to several examined downstream tasks.
We evaluate our generated test cases on the three state-of-the-art code embedding methods widely used in software engineering community, {\it i.e.}, Code2vec \cite{alon2019Code2vec}, Code2seq \cite{alon2018Code2seq} and CodeBERT \cite{feng2020codebert}, each with four different downstream tasks.

The contributions of this paper are summarized as follows:

\begin{itemize}[leftmargin=*]
    \item Proposing a search-based testing framework for adversarial robustness testing of code embedding models.
    \item Improving the robustness of code embedding models by retraining the models using the generated adversarial examples.
    \item Empirical evaluation of the proposed approach on three state-of-the-art embedding models, each with four downstream tasks.
\end{itemize}

\section{Background}
In this section, we briefly discuss the background on DNN testing, code embedding, and adversarial sample generation.

\subsection{DNN Testing}
We introduce the existing testing techniques for DNN, especially on the testing criteria, including neuron coverage and mutation score. More comprehensive discussion can be referred to the recent survey~\cite{zhang2020machine}.

\subsubsection{DNN Structural Coverage}

In traditional software testing, code coverage ({\it e.g.}, statements, branches) is a widely used indicator to measure testing sufficiency~\cite{li2017transforming}. 
Unlike traditional software, deep learning does not implement program logic through explicit statements, branches, and conditions. Instead, the logic is automatically learned and encoded into the neural network\cite{pei2017deepxplore}, following the data-driven programming style. Thus recent works on DNN testing have introduced multiple DNN structural coverage criteria based on ``neuron'' activation status to assess how well the input data have covered the runtime states of a DNN \cite{pei2017deepxplore, ma2018deepgauge,swrhkk2018, ma2019deepct}.
In particular, Pei {\it et al.} \cite{pei2017deepxplore} first introduced Neuron Coverage (NC), as a testing metric for DNNs. NC is defined as the ratio of the activated neurons for the given test inputs to the total number of neurons in the DNN model. Inspired by NC, many DNN structural coverage based on single neuron and neuron interactions are proposed ({\it e.g.}, DeepGauge \cite{ma2018deepgauge}, DeepConcolic, DeepCT \cite{swrhkk2018} \cite{ma2018deepgauge}).
The similar idea was also extended to stateful machine learning models such as recurrent neural networks~\cite{10.1145/3338906.3338954, 9286028}.
In addition, distribution and uncertainty based methods are also investigated~\cite{kim2019guiding,10.1145/3377811.3380368, 9286113}. For example, Kim {\it et al.}\cite{kim2019guiding} introduced the feature distribution-based criteria, called Surprise Coverage, which can estimate how an input  surprise the DNN.

\subsubsection{Mutation Testing}

Mutation testing~\cite{PAPADAKIS2019275} follows a white-box testing approach that modifies the original program with small changes \cite{budd1985program}. The mutated program ({\it i.e.}, mutant) are used to analyze whether test cases can detect the behavior change, comparing with the original program. Mutation score, defined as the ratio of detected mutants against all the seeded mutants, is often used to measure the quality of the test cases.

Ma {\it et al.} \cite{ma2018deepmutation,8952248} later introduced the mutation into the context of DNN, and proposed DeepMutation that mutates DNN models at the source-code or model-level, to make minor perturbations on the decision boundary of a DNN. They also define a mutation score as the ratio of test cases that their results are changed on the mutant versus the original program, over the total number of test cases. 
Compared with traditional software, the major difference in DNN mutation testing is the new mutation operator definition for DNN, which makes minor changes to the DNN decision logic.
With the generated mutant DNN models, the approximation of the DNN model robustness can be analyzed by the inference consistencies of the original DNN and mutant DNNs.
In particular, suppose a k-classification problem and let $C =$ \{$c_1,...,c_k$\} be all the $k$ classes of input data. For a test data point $t^{\prime} \in T^{\prime}$, $t^{\prime}$ kills $c_i \in C$ of mutant $m^{\prime} \in M^{\prime}$ if the following conditions are satisfied: (1) $t^{\prime}$ is correctly classified as $c_i$ by the original DL model $M$, and (2) $t^{\prime}$ is not classified as $c_i$ by $m^{\prime}$. Mutation score for DL systems is defined as follows:

\begin{equation}
    MutationScore(T^{\prime} , M^{\prime}) = \frac{\Sigma_{m^{\prime} \in M^{\prime}} |KilledClasses(T^{\prime},m^{\prime})|}{|M^{\prime}| \times |C^{\prime}|}
\end{equation}

where $KilledClasses(T^{\prime},m^{\prime})$ is the set of classes of $m^{\prime}$ killed by test data in $T^{\prime}$\cite{ma2018deepmutation}.

Wang {\it et al.} \cite{wang2019adversarial} propose an approach to detect adversarial samples using model mutation. Their approach is an integration of \textit{DeepMutation} testing \cite{ma2018deepmutation} and statistical hypothesis testing\cite{wald1947sequential}, which define the problem as how efficiently the model can decide whether $f(x)$ is a normal sample or an adversarial sample, given an input $x$ and a DNN model $f$. 
Their analysis is based on ``sensitivity'', which is measured by Label Change Rate (LCR). The assumption is the mutated DNN models are more likely to label an adversarial sample differently (compared to the label generated by the original DNN model). 
Given an input sample $x$ (either regular or adversarial data) and a DNN model $f$, DeepMutation first mutates the model using a set of model mutation operators, to create a set of mutated models $F$. Next, the label $f_i(x)$ of the input sample $x$ on every mutated model $f_i \in F$ is predicted. Finally, LCR is defined on a sample $x$, as follows:
\begin{equation}
    LCR(x) = \frac{|\{f_i|f_i \in F   ,   f_i(x) \ne f(x)\}|}{|F|}
\end{equation}

Intuitively, $LCR(x)$ measures how sensitive an input sample $x$ is on a DNN model's mutations, based on which an adversarial example is determined.

\subsection{Code Embedding}
\label{code_embedding}

Overall, Chen and Monperrus {\it et al.} \cite{chen2019literature} classify embeddings on source code into five categories: embedding of tokens, expressions, APIs, methods, and other miscellaneous embeddings. For instance, White {\it et al.} \cite{white2019sorting} used embedding of tokens in automatic program repair.  Alon {\it et al.} \cite{alon2019Code2vec} define embedding of functions using a notion of a path in the Abstract Syntax Tree (AST) of a Java method. Nguyen {\it et al.} \cite{nguyen2016mapping} use the embedding of sequences by applying embeddings on Java and C\# APIs to find similar API usage between the languages. Xu {\it et al.} \cite{xu2017neural} apply embedding of binary code on control flow graphs extracted from binary files.

Currently, there are three well-known embedding methods that stand out with public tool support, which have been extensively adopted by other researchers downstream source code processing tasks.

\begin{itemize}[leftmargin=*]
    \item \textbf{Code2vec} \cite{alon2019Code2vec} presents a neural model for encoding code snippets as continuously distributed vectors. It evaluates the embedding by predicting a given method name based on its body's source code, as a downstream task. Code2vec models the source code as AST paths. An AST path is defined as a path between nodes in the AST, starting from one terminal, ending in another terminal, and passing through an intermediate non-terminal node (a common ancestor of both terminals), in the path. Both source and destination terminals, along with the AST path, are mapped into an embedding vector, which is learned jointly with other networks during training. Each terminal and the path is then concatenated to a single context vector called path-context, which is also an attention vector to score each path-context to a single code vector, representing the method body.
    
    \item \textbf{Code2seq}  \cite{alon2018Code2seq} adopts an encoder-decoder architecture to encode paths node-by-node and creates labels as sequences, at each step. Similar to Code2vec, Code2seq uses a method prediction task for evaluation. The encoder represents a method body as a set of AST paths where each path is compressed to a fixed-length vector using a bi-directional LSTM, which encodes paths node-by-node. The decoder uses attention to select relative paths while decoding and predicts sub-tokens of target sequence at each step when generating the method name. This is different than Code2vec that uses monolithic path embeddings and only generates a single label at a time. 
    
    \item \textbf{CodeBERT} \cite{feng2020codebert} also learns general-purpose representations that support downstream software engineering tasks such as code search and code documentation generation. It is a \textit{bi-modal} pre-trained model for natural language (NL) and programming language (PL) like Python and Java. 

\end{itemize}

\subsection{Code Adversarial Models}
The most related work to our study are two code adversarial generator strategies ({\it i.e.}, \emph{1-Time} and \emph{K-Times} Mutation) that we explain them in this section and use them as comparison baseline in our experiment section. Both these techniques are based on refactoring source code, which is called mutation in this context (don't be confused with mutation in the context of evolutionary algorithms).

\subsubsection{1-Time Mutation}
1-Time Mutation method is a simple random refactoring method. In first analyzes all the Java source code to extract all the method code fragments that can be the target for mutation. Then, for each Java method, a refactor/mutator is randomly selected, from a pool of predefined refactoring operator, for mutation.
Note that some of the randomly selected refactoring operations might not be applicable to a particular Java method. For instance, if the specific method does not contain any loop, the randomly chosen ``Loop Enhance method'' cannot be applied there. Therefore, we iterate the process until we make sure that the method is refactored once.

\begin{figure}
\centering
\subfigure[A sample code snippet from Java-Large dataset. This code snippet can be refactored by 1-time, K-times and GM refactoring.]{
\begin{minipage}[b]{0.46\textwidth}
\includegraphics[width=1\textwidth]{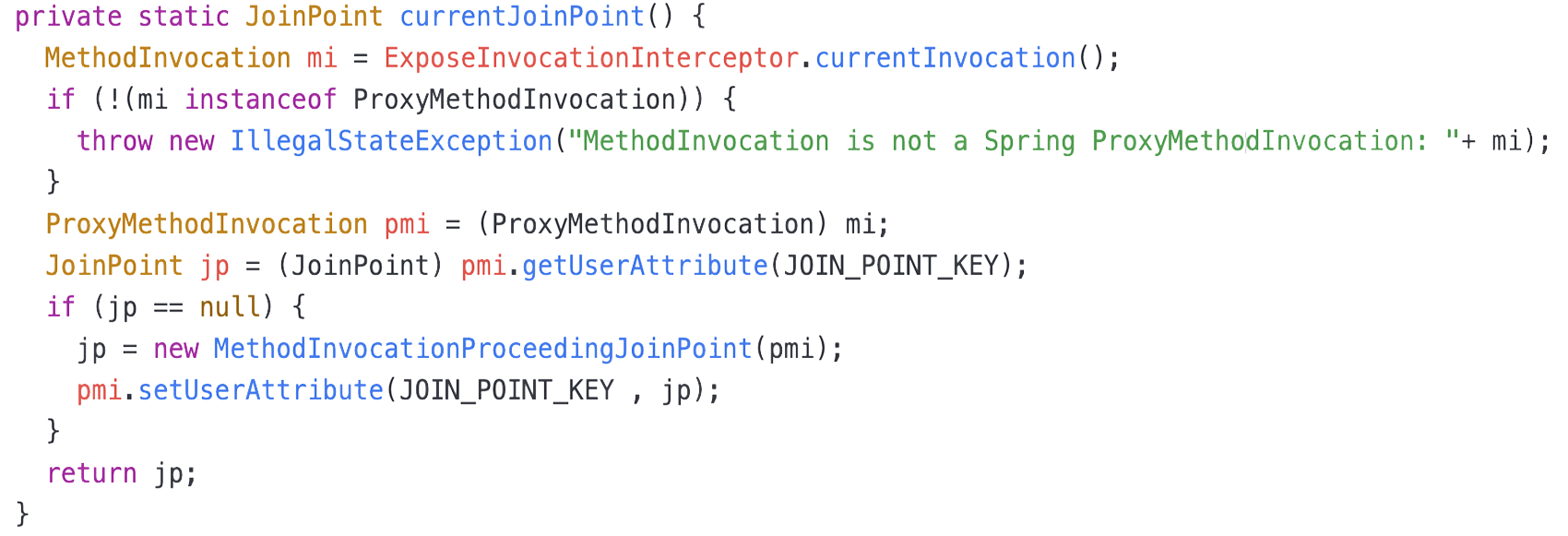}
\end{minipage}
}
\subfigure[Generated adversarial sample based on the example from (a) is from 1-Time refactoring method. The underlined line indicates the change from the original code snippet.]{
\begin{minipage}[b]{0.46\textwidth}
\includegraphics[width=1\textwidth]{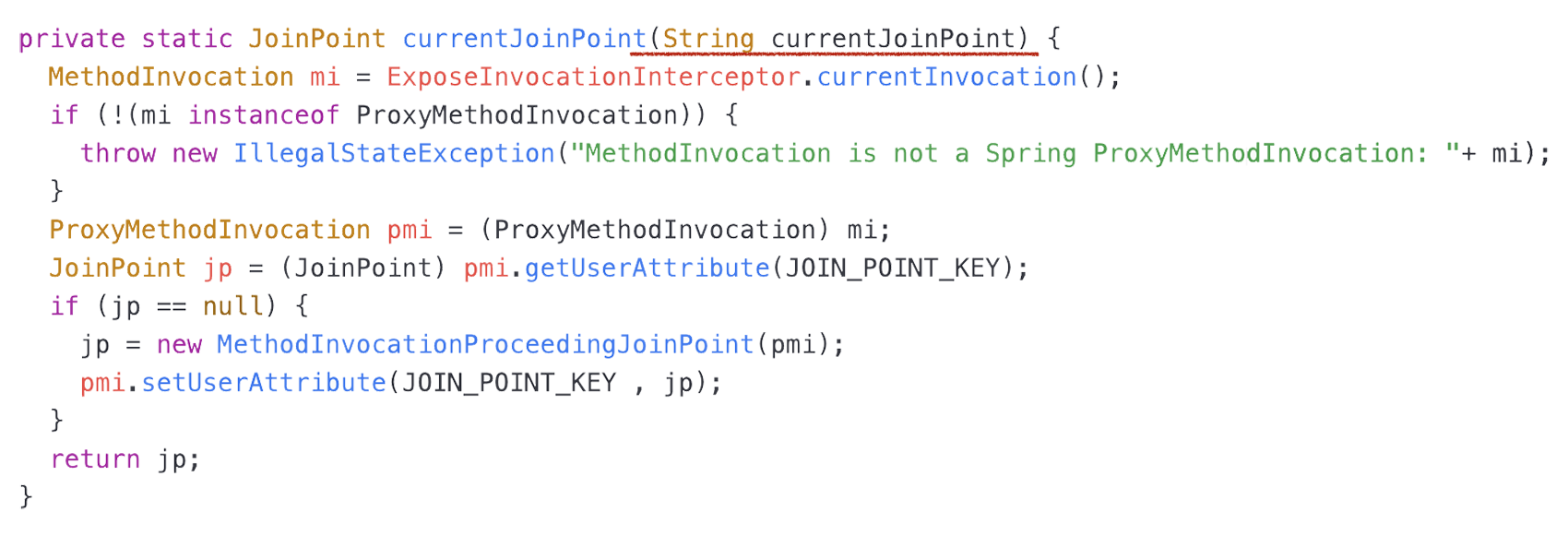}
\end{minipage}
}
\subfigure[Generated adversarial sample based on the example from (a), using 5-Times refactoring method. The underlined lines indicate the changes from original code snippet.]{
\begin{minipage}[b]{0.46\textwidth}
\includegraphics[width=1\textwidth]{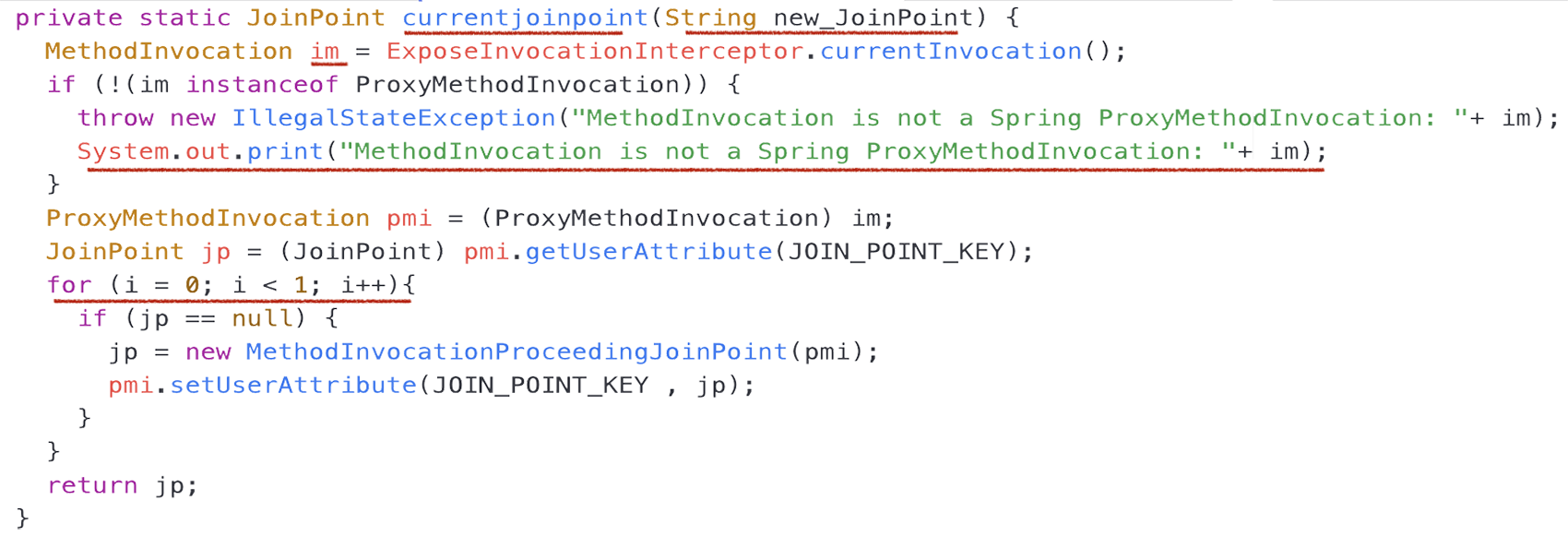}
\end{minipage}
}
\subfigure[Generated adversarial sample based on the example from(a), using Guided Mutation method. The underlined lines indicate the changes from original code snippet.]{
\begin{minipage}[b]{0.46\textwidth}
\includegraphics[width=1\textwidth]{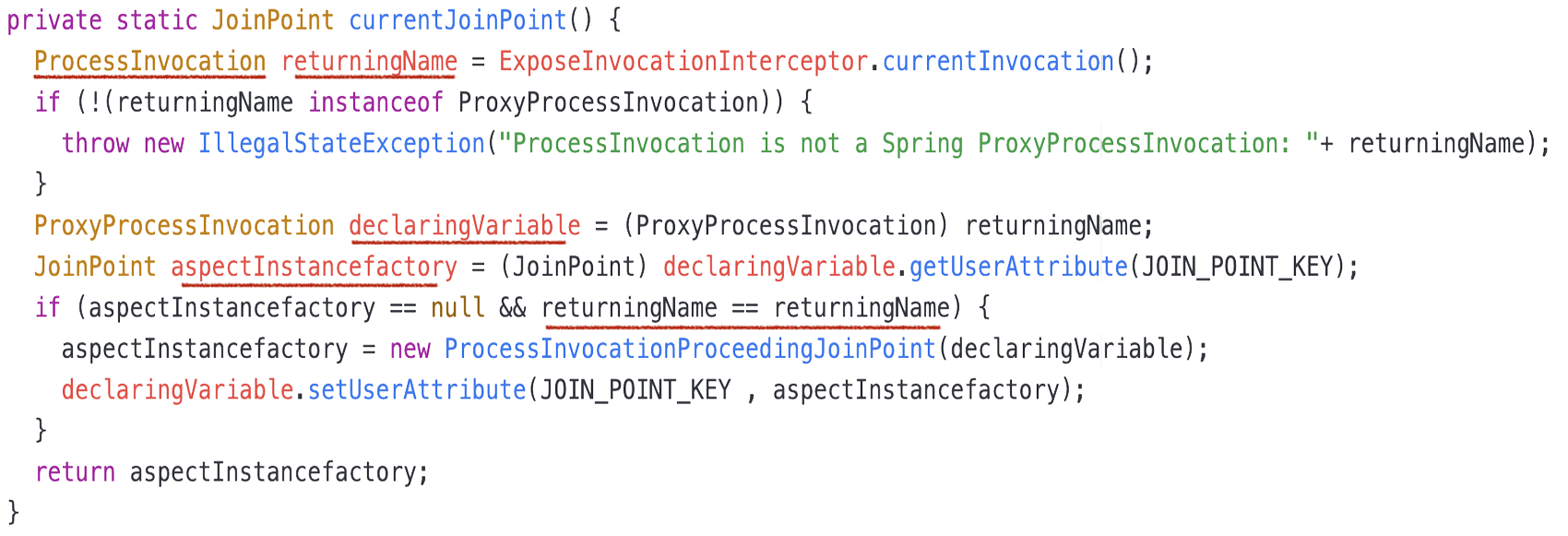}
\end{minipage}
}
\caption{Examples for the refactored code snippet.}
\label{refactored}
\end{figure}

Once all methods are extracted and refactored, the adversarial Java files are generated. Figure \ref{refactored}-(b) is a 1-Time refactored sample, create by applying ``Argument Adding'' refactoring operator on the code snippet from Figure \ref{refactored}-(a).


\subsubsection{K-Time Mutation}
\begin{figure}[t]
\centering
\includegraphics[width=60mm]{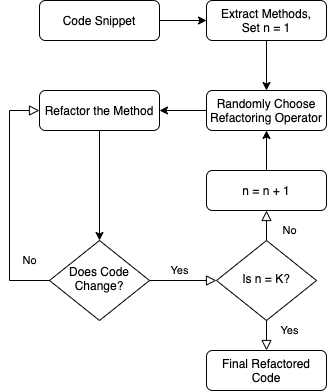}
\caption{The K-Times Mutation process for code adversarial generation, with K=5 \cite{ramakrishnan2020semantic}. The 1-Time can be seen as a variant of this approach with K=1.}
\label{5t_overall}
\end{figure}

Similar to the 1-time approach, K-times approach also performs random refactoring mutation on Java methods, except that the refactoring operation is performed K times.
In particular, after extracting each Java method, a randomly selected refactoring operator is applied, and this process would be repeated for K times, per method (see Figure \ref{5t_overall}). Again, some of the random refactoring operators might not be applicable to a given method. Therefore, we iterate the process with different operators to make sure the method is refactored K times. In this paper, we use K = 5, following the original study by Ramakrishnan {\it et al.} \cite{ramakrishnan2020semantic} that suggested K = 5 has the best F1 score of the test inputs. Figure \ref{refactored}-(c) shows a code snippet example, generated by the 5-Times adversarial technique. ``Local Variable Renaming'', ``Argument Adding'', ``For Loop Enhance'', ``Add Print'', and ``Method Name Renaming'' refactoring operators are the five random operators that are chosen in this example.

\section{Methodology}
\label{method}

Although extensively adversarial attack and testing studies have been performed, which show that the state-of-the-art DNNs still suffer robustness issues in the context of image, text and speech processing~\cite{chakraborty2018adversarial}, limited studies have been performed for DNNs in the context of source code processing \cite{zhang2019generating}.
Although sharing the discreteness characteristic with natural language, source code must strictly follow formal language grammar. Therefore, a general-purpose adversarial strategy often does not work, since the generated adversarial examples 
are often invalid and uncompilable.
For robustness testing of source code DNNs, code refactoring that transforms the code into the semantically equivalent form ({\it i.e.}, code snippet perturbation) can be a promising way. In particular, a robust DNN model should have the same inference results on the refactored code and its original counterpart, which forms the core test generation element of our search-based testing.
All refactoring methods and source code can be found here\footnote{\url{https://github.com/MaryamVP/Guided-Mutation-ICST-2021}}.

\subsection{Refactoring as the Test Generation Basis} 
In software engineering, code refactoring is a way to change a code snippet while keeping the original semantics as much as possible. It is used to improve the existing code by making it more readable, understandable, and clean. Refactoring also helps to add new features, reform the functions, building large applications easier, and detecting bugs faster.

In this paper, we adopt refactoring operators to generate adversarial samples for source code. 
To be comprehensive, we select ten popular refactoring operators widely used in Java, including those used in the previous work \cite{ramakrishnan2020semantic, rabin2020evaluation}:

\begin{itemize}[leftmargin=*]
    \item \textbf{Local Variable Renaming}: Renames the name of a variable using synonym words ({\it e.g.}, LIST and ARRAY, INSERT and ADD, FIND and SEARCH, {\it etc.}).
    \item \textbf{Argument Renaming}: Renames the name of an argument using a synonym word.
    \item \textbf{Method Name Renaming}: Renames the name of a method using a synonym word.
    \item \textbf{API Renaming}: Renaming the name of an API by using a synonym word for the local variable. API parameters determine the type of action one wants to take on the resource. Each parameter has a name, value type and optional description. Renaming the API can create the refactored codes, with similar functionality.
    \item \textbf{Local Variable Adding}: Adds a local variable to the code.
    \item \textbf{Argument Adding}: Adds an argument to the code.
    \item \textbf{Add Print}: Adds print to a random line of the code.
    \item \textbf{For Loop Enhance}: Replaces \textit{for} loops with \textit{while} loops or vice versa.
    \item \textbf{IF Loop Enhance}: Replaces an IF condition with an equivalent logic. 
    \item \textbf{Return Optimal}: Changes a return variable where possible.
\end{itemize}

Given that the source code functionality has not been changed using the above-mentioned refactoring operators, if the DNN result changes, we call the refactored code as an adversarial sample, which triggers the robustness issue of the DNN. For the sake of terminology consistency with previous studies, we call the refactoring as the mutation and a refactored Java file as a mutant.

\begin{table}[t]
\centering
\scriptsize
\setlength\tabcolsep{3.2pt}
\caption{Mutation Operators}
\begin{center}
\begin{tabular}{|c|c|c|c|}
\hline
\multicolumn{2}{|c|}{\textbf{Level}} & \textbf{Operator}  & \textbf{Description}\\ \hline
\multirow{2}{*}{\textbf{\textbf{Static}}}  & \multirow{2}{*}{\textbf{Weight}} & Weight Gaussian Fuzzing (WGF)    & Fuzz weights         \\ \cline{3-4} 
&     & Weight Precision Reduction (WPR) & Reduce weight precision      \\ \hline
\multirow{7}{*}{\textbf{Dynamic}} & \multirow{4}{*}{\textbf{State}}  & State Clear (SC)       & Clear the state to 0 \\ \cline{3-4} 
     &     & State Reset (SR)       & Reset state to previous state  \\ \cline{3-4} 
     &     & State Gaussian Fuzzing (SGF)     & Fuzz state value     \\ \cline{3-4} 
     &     & State Precision Reduction (SPR)  & Reduce state value precision \\ \cline{2-4} 
     & \multirow{3}{*}{\textbf{Gate}}   & Gate Clear (GC)        & Clear the gate value to 0      \\ \cline{3-4} 
     &     & Gate Gaussian Fuzzing (GGF)      & Fuzz gate value      \\ \cline{3-4} 
     &     & Gate Precision Reduction (GPR)   & Reduce gate value precision  \\ \hline
\end{tabular}
\label{operators}
\end{center}
\end{table}

\subsection{Guided Mutation: A Search-based Testing Framework}

To guide effective DNN testing in the huge testing space, in this section, we propose a  A Search-based Testing Framework guided by mutation testing ({\it i.e.} GM).

GM adopts an evolutionary strategy and follows the same workflow as GA, except that we only apply mutation but not crossover operations on the input population. 
The reason is that changing the code snippets using crossover may cause many incorrect (not even compilable) code snippets, letting alone functionality preserving code. Recall that the goal of an adversarial sample is to be as similar to the original data (so that the perturbation is minor). Therefore, in the code embedding domain, if the generated adversarial samples are throwing run-time or compile-time errors, they are too easy to be called adversarial samples.
That is why we did not include crossover and define mutations based on refactoring operators which guarantee semantic preservation. 
In our framework GM, we adopt Elitism that involves copying a small portion of the fittest candidates, unchanged, into the next generation. It can sometimes have a dramatic impact on performance by ensuring that the GM does not waste time re-discovering previously discarded partial solutions. Candidate solutions that are preserved unchanged through elitism remain eligible for selection as parents when developing the next generation's remainder.

As shown in Figure \ref{gm_flow}, the concrete steps to generate adversarial samples by GM are as follows:

\begin{figure}[t]
\centering
\includegraphics[width=90mm]{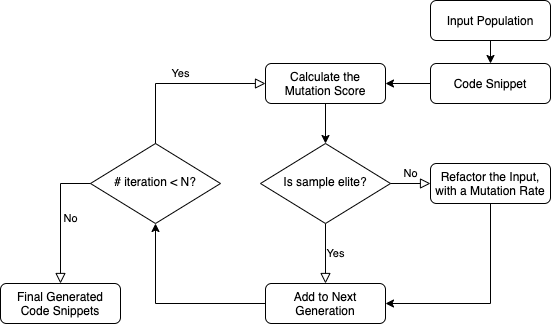}
\caption{The guided mutation (GM) process for code adversarial generation.}
\label{gm_flow}
\end{figure}

\begin{enumerate}[leftmargin=*]
    \item Calculate mutation score for the current population of code snippets.
    \item Choose elite candidates based on the highest mutation score and copy them into the next generation's population.
    \item Mutate the remaining candidates with specified mutation rates.
    \item Repeat from step one until meeting the stopping criterion ({\it e.g.}, reaching a specified number of iterations).
\end{enumerate}

The only remaining part of the algorithm to discuss is the mutation score. In this work we adopt DeepMutation++ \cite{hu2019deepmutation++} to calculate the mutation score per code snippet, the DNN mutators of which are summarized in  Table \ref{operators}

Given an input $t$, a DNN $m$ and its mutant $m^\prime$, they say $t$ is killed by $m^\prime$ if the outputs are inconsistent at $t$, {\it i.e.}, $m(t) \neq m^\prime(t) $. Given a set of mutant DNNs, $M$, they define the mutation score as:

\begin{equation}
    MS(t,m,M) = \frac{|\{m^\prime| m^\prime \in M \wedge m(t) \neq m^\prime(t)\}|}{|M|}
\end{equation}

In this paper, we used the mutation score as our fitness function for the GM model. We use all nine operators for the RNN models and created ten refactored Java codes using each operator. Therefore, overall we have 90 refactored files for a model. Figure \ref{refactored}-(d) shows sample adversarial code examples generated GM and the specific refactors contain \textit{IF Loop Enhance}, \textit{Local Variable Renaming}, {\it etc.}


\subsection{Retraining Procedure for robustness improvement} 

So far, we explained three methods for generating adversarial code examples: 1-time, K-times and GM. In this paper, we have two high-level objectives that are (a) robustness testing of code embedding models and (b) improving the robustness of the downstream tasks. For the first objective, we generate adversarial data based on the original test set. The idea is to test the robustness of the trained model (trained on the original train set). To study the second objective, we retrain the code embedding model using an augmented train set which includes the original train set plus new adversarial examples that are created by applying an adversarial code generator on the original train set. Retraining here refers to re-running the process that generated the previously selected model on a new data training set. The features, model algorithm, and hyper-parameter search space should all remain the same. 
The intuition is that such an augmented train set can potentially help to improve the robustness of the model, which increases its performance on the final downstream task.

\section{Empirical Evaluation}
In this section, we perform comprehensive evaluations on three code embedding models and four downstream tasks, to evaluate our robustness testing and improvement methods.

\subsection{Objectives and research questions}
The objectives of this study are (a) to evaluate our proposed code adversarial generation strategy to see how effective it is in testing the robustness of code embedding models, and (b) to assess the improvement of some downstream tasks when their underlying code embedding models are retrained by our robustness improvement strategy. 

In particular, we investigate the following research questions:
\begin{itemize}[leftmargin=*]
    \item \textbf{RQ1:} How robust are code embedding models with respect to adversarial code samples?
    \item \textbf{RQ2:} How much does re-training the embedding model with adversarial samples improve the robustness and performance of the downstream tasks?
\end{itemize}

\subsection{DNN models and configurations}
Our evaluation selects three state-of-the-art and widely used embedding models for code, i.e., Code2vec \cite{alon2019Code2vec}, Code2seq \cite{alon2018Code2seq} and CodeBERT \cite{feng2020codebert}, which are all publicly available to replicate. Similar to Rabin {\it et al.} \cite{rabin2020evaluation}, we set the number of epochs as 20 (no more significant improvement was seen after that) and kept other configurations as suggested in the original code embedding papers. For the GM experiment, we set the mutation rate as 0.05,  which is suggested to be a suitable mutation rate for a genetic algorithm in previous studies \cite{haupt2000optimum}.

\subsection{Dataset}
In this paper, we used the original datasets provided by the embedding tools. Although Code2seq and CodeBERT have publicly published their input datasets, Code2vec only made pre-processed files accessible to the public. Fortunately, the structure of preprocessing files for both Code2vec and Code2seq was similar. Therefore, we used the original dataset that comes with Code2seq for the Code2vec model, as well.


The dataset used for Code2vec and Code2seq is called ``Java-Large dataset'', which is available in the Code2seq GitHub page\footnote{\url{https://github.com/tech-srl/Code2seq}}. It includes 9,000 Java projects for training, 200 for validation, and 300 for testing. This dataset contains about 16M examples. Note that since Code2vec splits the dataset by a single Java \textit{file} but not Java \textit{projects}, we organized all the Java files in a single folder. Overall, we had about 1.9M Java files for Code2vec and Code2seq experiments, whose size is about 20GB in total.

CodeBERT also use a recent large dataset provided by Husain {\it et al.} \cite{husain2019codesearchnet}, which includes
2.1M bimodal datapoints and 6.4M unimodal codes across six programming languages (Python, Java, JavaScript, PHP, Ruby, and Go), as suggested for default parameters, available in the CodeBERT website\footnote{\url{https://github.com/microsoft/CodeBERT}}.

Code2vec and Code2seq both support Java and C\# languages, while Code2seq supports Python language, as well. Despite the large language support for these tools, we only experimented with Java language code snippets since it was the only common language among all models. 

After having all the original files, we applied applicable refactoring methods to the original Java files in both training and testing datasets. The total number of generated programs is 1,798,419 training files, 44,140 validation files, and 59,404 test files for each experiment.

In this paper, we use 16 different experiments to evaluate our generated adversarial examples, per embedding and task. These 16 experiments include four different sets of test data and four different sets of train data. For the training datasets, we have the four following sets:

\begin{itemize}[leftmargin=*]
    \item Original training dataset (that comes with the original tool).
    \item Original training dataset + adversarial samples generated by applying 1-Time refactoring method on the training dataset.
    \item Original training dataset + adversarial samples generated by applying 5-Times refactoring method on the training dataset.
    \item Original training dataset + adversarial samples generated by applying GM refactoring method on the training dataset.
\end{itemize}

We also prepared four test sets as follows:

\begin{itemize}[leftmargin=*]
    \item Original test dataset.
    \item Adversarial samples generated by applying 1-Time on original test dataset (only refactored codes not the original ones).
    \item Adversarial samples generated by applying 5-Times on original test dataset (only refactored codes not the original ones).
    \item Adversarial samples generated by applying GM on original test dataset (only refactored codes not the original ones).
\end{itemize}

Then, for each embedding and downstream task at hand, we train the models on one of the four training sets and test it on one of the test datasets (Total=4*4=16).  

% ...

\subsection{Evaluation Metrics}
Next, we give a brief description of the evaluation metrics used in different downstream tasks in our experiments: 
\begin{itemize}[leftmargin=*]
    \item \textbf{F1-Score}: F1-score is a measure of classification accuracy. It is calculated from the precision and recall of the test. The precision is the number of correctly identified positive results divided by the number of all positive results, including those not identified correctly. The recall is the number of correctly identified positive results divided by the number of all samples that should have been identified as positive. F1-score is calculated as below:
    \begin{equation}
        F1 = 2 * \frac{Precision * recall}{precision + recall}
    \end{equation}
    This metric has been reported by all three models on the following tasks (on method name prediction in both Code2vec and Code2seq and on coed search in CodeBERT).
    
    \item \textbf{ROUGE}: ROUGE, or Recall-Oriented Understudy for Gisting Evaluation\cite{lin2004rouge} is a set of metrics and a software package used for evaluating automatic summarization and machine translation software in natural language processing. In this paper, it is used for Code Captioning task. The metrics compare an automatically produced summary or translation against a human-produced summary or translation. The following evaluation metrics have been used in this study:
    
    \begin{itemize}[leftmargin=*]
        \item \textbf{ROUGE-N:} The overlap of N-grams \cite{lin2003automatic} between the system and reference summaries. For example, \textit{ROUGE-1} refers to the overlap of unigram (each word) between the system and reference summaries; \textit{ROUGE-2} refers to the overlap of bigrams between the system and reference summaries.
        \item \textbf{ROUGE-L:} The Longest Common Subsequence (LCS) based statistics. Longest common subsequence problem takes into account sentence-level structure similarity, naturally, and identifies longest co-occurring in sequence n-grams, automatically.
    \end{itemize}
    
    \item \textbf{BLEU}: BLEU (bilingual evaluation understudy) is an algorithm for evaluating the quality of a machine-translated text from the text. Quality is considered the correspondence between the output of a machine and a professional human translation. It has been used with CodeBERT and the Code Document Generation Task \cite{papineni2002bleu} in this paper.
\end{itemize}

\subsection{Downstream Source Code Processing Tasks}
\label{downstream_tasks}
In this study, we evaluate the trained model on four different downstream tasks: Method Name Prediction, Code Captioning, Code Search and Code Documentation Generation.

\begin{itemize}[leftmargin=*]
    \item \textit{\textbf{Method Name Prediction}}: Predict method name given the method body. The evaluation metric is F1-score over sub-tokens.
    
    \item \textit{\textbf{Code Captioning}}: Predict a full natural language sentence given a short Java code snippet. The target sequence length in this task is about ten on average. The model is evaluated using ROUGE-N and ROUGE-L F1-Score.

    \item \textit{\textbf{Code Search}}: Given a natural language as the input, the objective of code search is to find the most semantically related code from a collection of codes. The evaluation metric is F1-score.
    
    \item \textit{\textbf{Code Documentation Generation}}: Generate software documentation intended for programmers (API documentation) or end-users (end-user guide), or both, from a set of source code files. The model is evaluated using the BLEU metric.

\end{itemize}

Table \ref{orig_res_all} summarizes the score of each model, as reported in the respective original paper, on their downstream tasks. While the trained models for Code2seq and CodeBERT are consistent with the performance reported in their original papers, our Code2vec could not reach the F1 score reported in the original paper, because raw dataset of the original paper was missing and we used Code2seq data here (which might not be exactly the same).
Note that Code2seq did not experiment Code Captioning on Java language, hence we did not add it to the table.

\begin{table}[]
\centering
\caption{The three models' F1-score reported in their original papers with default configurations on default dataset.}
\begin{tabular}{|c|c|c|c|}
\hline
\textbf{\begin{tabular}[c]{@{}c@{}}Code Embedding \\ Model\end{tabular}}   & \textbf{\begin{tabular}[c]{@{}c@{}}Evaluation \\ Metric\end{tabular}} & \textbf{Score} & \textbf{Downstream Task} \\ \hline
Code2vec  & F1 Score & 58.40 & Method Name Prediction \\ \hline
Code2seq & F1 Score & 59.19 & Method Name Prediction \\ \hline
\multirow{2}{*}{CodeBERT} & F1 Score 
& 74.84 & Code Search \\ \cline{2-4} 
& BLEU   & 0.79 & Code Document Generation \\ \hline
\end{tabular}
\label{orig_res_all}
\end{table}

\begin{table*}[t]
\centering
\caption{The performance of a model-task combination on the original and adversarial test sets, generated using 1-Time, 5-Times and GM techniques. In this table, the ``Improvement\%'' (the values in parenthesis) is the model score on the adversarial test sets minus its score on the original test set divided by the original test set score, per model-task combination.}
\begin{tabular}{|c|c|c|c|c|c|c|}
\hline
\multirow{2}{*}{\textbf{\begin{tabular}[c]{@{}c@{}}Embedding\\ Model\end{tabular}}} & \multirow{2}{*}{\textbf{\begin{tabular}[c]{@{}c@{}}Downstream\\ Task\end{tabular}}} & \multirow{2}{*}{\textbf{Evaluation Metric}}& \multicolumn{4}{c|}{\textbf{Test Set}}       \\ \cline{4-7} 
 &   &  & \textbf{Original} & \textbf{\begin{tabular}[c]{@{}c@{}}1-Time Adversarial\\ (Improvement\%)\end{tabular}} & \textbf{\begin{tabular}[c]{@{}c@{}}5-Times Adversarial\\ (Improvement\%)\end{tabular}} & \textbf{\begin{tabular}[c]{@{}c@{}}GM Adversarial\\ (Improvement\%)\end{tabular}} \\ \hline
Code2vec      & Method Name Prediction & F1 Score   & 35.92   & 34.75 (-3.26\%) & 33.87 (-5.71\%) & 35.45 (-1.31\%)         \\ \hline
\multirow{2}{*}{Code2seq}  & Method Name Prediction & F1 Score   & 42.71   & 40.78 (-4.52\%)& 40.16 (-5.97\%)& 38.66 (-9.48\%)         \\ \cline{2-7} 
 & Code Captioning& \begin{tabular}[c]{@{}c@{}}ROUGE\\ F1 Score\end{tabular} & 52.09   & 43.16 (-17.14\%)& 47.55 (-8.72\%)& 49.36 (-5.24\%)         \\ \hline
\multirow{2}{*}{CodeBERT}  & Code Search    & F1 Score   & 81.36   & 80.19 (-1.44\%) & 79.81 (-1.91\%)& 80.65 (-0.87\%)         \\ \cline{2-7} 
 & Code Document Generation & BLEU       & 0.79    & 0.62 (-21.52\%) & 0.65 (-17.72\%) & 0.71 (-10.13\%)\\ \hline
\multicolumn{4}{|c|}{\textbf{Mean of Improvement Percentages}}& \textbf{-9.58\%}    & \textbf{-8.01\%}    & \textbf{-5.41\%}\\ \hline
\multicolumn{4}{|c|}{\textbf{Median of Improvement Percentages}}     & \textbf{-4.52\%}    & \textbf{-5.97\%}    & \textbf{-5.24\%}\\ \hline
\end{tabular}
\label{rq1_table}
\end{table*}

\subsection{Execution environment}
We run 16 experiments per model and downstream task (overall 80 different experiments, since we had 5 combinations of embedding/tasks). For our experiments, we use a cluster with 32 nodes, 24 cores, 250G memory, 2 x Intel E5-2650 v4 Broadwell @ 2.2GHz CPU, 1 x 800G SSD storage and 4 x NVIDIA P100 Pascal (16G HBM2 memory) GPU. CodeBERT takes three days, and Code2vec and Code2seq take four days to train using their original dataset and default configurations. Running 1-Time and K-Times to generate adversarial examples takes 5 hours while running GM algorithms on all three models takes 30 hours (on average 10 hours per model, which is 2X more expensive than baselines). However, the robustness improvement procedure (retraining), takes longer than original training since the number of input examples is doubled (original + adversarials). Therefore, our retraining on CodeBERT took four days, on Code2vec took six days, and on Code2seq took eight days, depending on the number of adversarial examples.

\subsection{Experimental Results}

This section demonstrates the performance of three adversarial test generation techniques to test DNN models' robustness for code embedding. In each experiment, an adversarial test generation technique was applied on a downstream task, for a code embedding model. In the rest of this section, we answer our two RQs:

\begin{table*}[t]
\centering
\caption{The performance of three adversarial code generation techniques (each augmenting the original training set) on 15 (5 embedding-tasks * 3 adversarial test sets) data sets. In this table, the ``Improvement\%'' (the values in parenthesis) is the retrained model's score minus the originally trained model score divided by the originally trained model score (all tested on 15 different datasets). }
\begin{tabular}{|c|c|c|c|c|c|c|c|}
\hline
\multirow{2}{*}{\textbf{\begin{tabular}[c]{@{}c@{}}Embedding\\ Model\end{tabular}}} & \multirow{2}{*}{\textbf{\begin{tabular}[c]{@{}c@{}}Downstream\\ Task\end{tabular}}} & \multirow{2}{*}{\textbf{Model Score}}& \multirow{2}{*}{\textbf{Test Set}} & \multicolumn{4}{c|}{\textbf{Training Set}}  \\ \cline{5-8} 
         &   &       &   & \textbf{Original} & \textbf{\begin{tabular}[c]{@{}c@{}}1-Time + original \\ (Improvement\%)\end{tabular}} & \textbf{\begin{tabular}[c]{@{}c@{}}5-Time + original \\ (Improvement\%)\end{tabular}} & \textbf{\begin{tabular}[c]{@{}c@{}}GM + original \\ (Improvement\%)\end{tabular}} \\ \hline
Code2vec & Method Name Prediction  & F1 Score     & 1-Time& 34.75   & 35.92 (3.37\%) & 35.48 (2.1\%) & 47.34 (36.23\%)     \\ \hline
\multirow{2}{*}{Code2seq}        & Method Name Prediction  & F1 Score     & 1-Time& 40.78   & 42.34 (3.83\%) & 43.29 (6.15\%) & 49.11 (20.43\%)      \\ \cline{2-8} 
         & Code Captioning       & \begin{tabular}[c]{@{}c@{}}ROUGE\\ F1 Score\end{tabular} & 1-Time& 43.16   & 43.39 (0.53\%)& 46.98 (8.85\%)& 49.82 (15.43\%)      \\ \hline
\multirow{2}{*}{CodeBERT}        & Code Search & F1 Score     & 1-Time& 80.19   & 81.45 (1.57\%)& 81.76 (1.96\%)& 82.37 (2.72\%)      \\ \cline{2-8} 
         & Code Document Generation  & BLEU  & 1-Time& 0.62    & 0.64 (3.23\%) & 0.65 (4.84\%)& 0.76    (22.58)   \\ \hline
Code2vec & Method Name Prediction  & F1 Score     & 5-Times& 33.87   & 35.79 (5.67\%)& 36.16 (6.76\%)& 46.51 (22.58\%)     \\ \hline
\multirow{2}{*}{Code2seq}        & Method Name Prediction  & F1 Score     & 5-Times& 40.16   & 42.35 (5.45\%)& 45.91 (14.32\%)& 52.22 (30.03\%)     \\ \cline{2-8} 
         & Code Captioning       & \begin{tabular}[c]{@{}c@{}}ROUGE\\ F1 Score\end{tabular} & 5-Times& 47.55   & 45.84 (-3.6\%)& 49.08 (3.22\%)& 55.12  (15.92\%)    \\ \hline
\multirow{2}{*}{CodeBERT}        & Code Search & F1 Score     & 5-Times& 79.81   & 80.36 (0.69\%)& 82.84 (3.8\%)& 82.52 (3.4\%)      \\ \cline{2-8} 
         & Code Document Generation  & BLEU  & 5-Times& 0.65    & 0.67 (3.08\%)& 0.70 (7.69\%)& 0.79 (21.54\%)      \\ \hline
Code2vec & Method Name Prediction  & F1 Score     & GM& 35.45   & 38.73 (9.25\%)& 37.08 (4.6\%)& 53.13 (49.87\%)      \\ \hline
\multirow{2}{*}{Code2seq}        & Method Name Prediction  & F1 Score     & GM& 38.66   & 43.89 (13.53\%)& 44.49 (15.08\%)& 56.24 (45.47\%)     \\ \cline{2-8} 
         & Code Captioning       & \begin{tabular}[c]{@{}c@{}}ROUGE\\ F1 Score\end{tabular} & GM& 49.36   & 46.96 (-4.86\%)& 54.28 (9.97\%)& 61.73 (25.06\%)     \\ \hline
\multirow{2}{*}{CodeBERT}        & Code Search & F1 Score     & GM& 80.65   & 82.33 (2.08\%)& 82.49 (2.28\%)& 82.97 (2.88\%)     \\ \cline{2-8} 
         & Code Document Generation  & BLEU  & GM& 0.71    & 0.73 (2.82\%)& 0.74 (4.23\%)& 0.83 (16.90\%)      \\ \hline
\multicolumn{5}{|c|}{\textbf{Mean of Improvement Percentages}}  & \textbf{3.11\%}  & \textbf{6.39\%}  & \textbf{
23.05\%}  \\ \hline
\multicolumn{5}{|c|}{\textbf{Median of Improvement Percentages}} & \textbf{3.08\%}  & \textbf{4.84\%}  & \textbf{21.54\%}  \\ \hline
\end{tabular}
\label{rq2_1_table}
\end{table*}

\begin{table*}[t]
\centering
\caption{The performance of retrained models on the original test sets. In this table, the ``Improvement\%'' (the values in parenthesis) is the retrained model's score minus the originally trained model score divided by the originally trained model score (all tested on the original test set). }
\begin{tabular}{|c|c|c|c|c|c|c|c|}
\hline
\multirow{2}{*}{\textbf{\begin{tabular}[c]{@{}c@{}}Embedding\\ Model\end{tabular}}} & \multirow{2}{*}{\textbf{\begin{tabular}[c]{@{}c@{}}Downstream\\ Task\end{tabular}}} & \multirow{2}{*}{\textbf{Model Score}}                    & \multirow{2}{*}{\textbf{Test Set}} & \multicolumn{4}{c|}{\textbf{Training Set}}                  \\ \cline{5-8} 
             &              &            &            & \textbf{Original} & \textbf{\begin{tabular}[c]{@{}c@{}}1-Time + original \\ (Improvement\%)\end{tabular}} & \textbf{\begin{tabular}[c]{@{}c@{}}5-Times + original \\ (Improvement\%)\end{tabular}} & \textbf{\begin{tabular}[c]{@{}c@{}}GM + original \\ (Improvement\%)\end{tabular}} \\ \hline
Code2vec     & Method Name Prediction              & F1 Score   & Original   & 35.92             & 36.48 (1.56\%)    & 36.4 (1.34\%)    & 36.14 (0.61\%)\\ \hline
\multirow{2}{*}{Code2seq}             & Method Name Prediction              & F1 Score   & Original   & 42.71             & 39.59 (-7.31\%)   & 39.98 (-6.39\%)    & 41.19 (-3.56)\\ \cline{2-8} 
             & Code Captioning        & \begin{tabular}[c]{@{}c@{}}ROUGE\\ F1 Score\end{tabular} & Original   & 52.09             & 40.91  (-21.46\%)  & 34.67  (-33.44\%)  & 36.19 (-30.52\%)\\ \hline
\multirow{2}{*}{CodeBERT}             & Code Search  & F1 Score   & Original   & 81.36             & 80.16  (-1.47\%)  & 80.53 (-1.02\%)   & 81.68 (0.39\%)\\ \cline{2-8} 
             & Code Document Generation              & BLEU       & Original   & 0.79              & 0.53   (-32.91\%)  & 0.68  (-13.92\%)   & 0.75 (-5.06\%) \\ \hline
\multicolumn{5}{|c|}{\textbf{Mean of Improvement Percentages}}                      & \textbf{-12.32\%}      & \textbf{-10.69\%}     & \textbf{-7.63\%} \\ \hline
\multicolumn{5}{|c|}{\textbf{Median of Improvement Percentages}}& \textbf{-7.31\%}     & \textbf{-6.39\%}    & \textbf{-3.56\%}\\ \hline
\end{tabular}
\label{rq2_2_table}
\end{table*}


\noindent\textbf{RQ1: How robust are code embedding models with respect to adversarial code samples?}

To answer this research question, we look at three adversarial code example generation techniques: 1-Time, 5-Times, and the GM method. We apply those techniques to the original test sets and create three new test sets per embedding-task. Each task uses its own evaluation metric (F1, Rouge, and BLEU). So the scores are not comparable across rows. However, we can look into the normalized differences between scores when the models are applied to the original test vs. the adversarial tests. These values in parentheses (``Improvement\%'') represent these normalized diffs and are now comparable across the rows. Table \ref{rq1_table} reports all scores and their ``Improvement\%''. It also summarizes the results as median and mean of Improvement Percentages.
The first observation is that, as expected, the model performance drops when applied on the adversarial test data compared to the original test set, since the whole idea of adversarial generation was to fool the models. 
We take an example to better understand how this happens.

\begin{figure}[t]
\centering
\includegraphics[width=50mm]{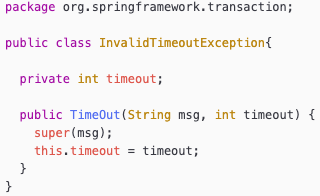}
\caption{Java code snippet example: \textit{InvalidTimeoutException.Java}}
\label{new_orig}
\end{figure}

\begin{figure}[t]
\centering
\includegraphics[width=50mm]{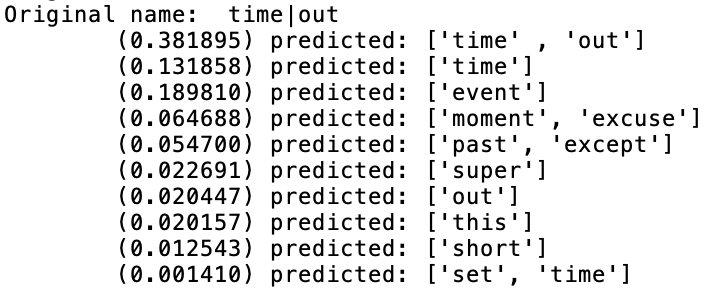}
\caption{Predicted method names and their probabilities for the original test inputs shown in Figure \ref{new_orig}.}
\label{orig_pred}
\end{figure}

In Figure \ref{new_orig} demonstrates a sample Java code snippet. Class \textit{InvalidTimeoutException} has a method called \textit{TimeOut}. Figure \ref{orig_pred} is the prediction result for Code2vec model on name prediction task. As shown in the prediction, the method name is predicted correctly with $0.38 \%$ probability. 

\begin{figure}[t]
\centering
\includegraphics[width=50mm]{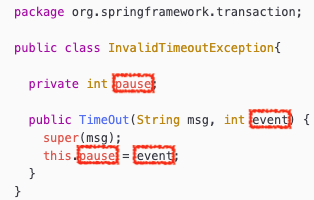}
\caption{GM refactored file for: \textit{InvalidTimeoutException.Java}}
\label{new_GM}
\end{figure}

\begin{figure}[t]
\centering
\includegraphics[width=50mm]{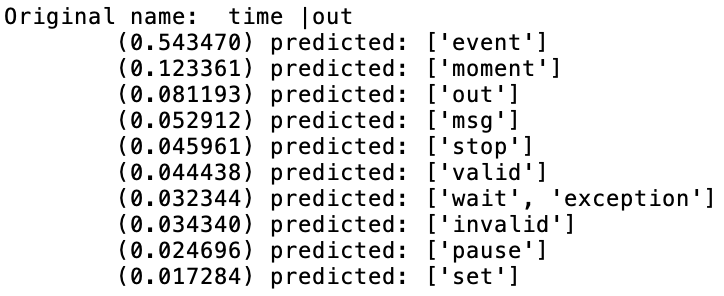}
\caption{Predicted method names and their probabilities for the GM test inputs shown in Figure \ref{new_GM}.}
\label{gm_pred}
\end{figure}

To confirm if generated adversarial examples can fool the model, we generated GM version of the \textit{InvalidTimeoutException.Java} file, shown in Figure \ref{new_GM}. As shown in Figure \ref{gm_pred}, the model could not predict the model name. Therefore, we can show that generating adversarials can fool the model, and therefore, decrease its F1-score.

The second observation is that the models are indeed relatively robust. Looking at the Median of Improvement Percentages (note that ``Improvement\%'' is basically a normalized diff between the model's score on the adversarial test set vs. the original test set), across all five embedding-tasks, show that the performance drops are relatively small, sitting at 4.52\% (for 1-Time), 5.97\% (for 5-Times), and 5.24\% (for GM). However, the mean values are a bit higher due to some outliers, especially for 1-Time adversarial.  Given that the medians are so close, we conclude that all three robustness testing approaches are equally effective in creating adversarial samples that negatively affect the code embedding-downstream task combinations, but overall impact is not large.  

Looking at the individual cases, however, we see that the ``Improvement\%'' values are between -0.87\% (for Code Search using CodeBERT tested on GM adversarials) to -21.52\% (for Code Document Generation using CodeBERT tested on 1-Time adversarials). The interesting observation is that even using the same embedding and the same adversarial test suite ({\it e.g.}, CodeBERT and 1-Time), two different tasks (Code Search and Code Document Generation) may result in very different drops in performance ({\it e.g.}, 1.44\% vs. 21.52\% on CodeBERT). This observation is not only for 1-Time. Similar patterns are also found for all three techniques. We can see that robustness of the downstream tasks may dominate the robustness of the embedding models and one should consider both together to best study the robustness.

\vspace{2mm}
\begin{tcolorbox}[size=title,colback=white]

{\textbf{Answer to RQ1}: 
Testing the original code embedding models using adversarial examples reduces the model's score (with a median normalized decrease of performance between 4.52\% and 5.24\%). The true robustness of embedding performance in the presence of adversarials, however, heavily depends on the downstream tasks. 
}
\end{tcolorbox}
\vspace{2mm}


\noindent\textbf{RQ2: How much re-training the embedding model with adversarial samples improves the robustness and performance of the downstream tasks?}

In this part, we answer our second research question by diving it into two sub-questions:

\noindent\textbf{RQ2.1. Which technique has the highest performance in terms of improving the robustness across all model-task combinations?}

To answer this sub-RQ, we look at the improvements of downstream tasks' scores when being retrained using an augmented training set. The augmented training sets include the original training data plus an equal-size adversarial dataset, generated by one of the three techniques of this study (1-Time, 5-Times, and GM). To evaluate the retrained models we use 15 different test sets. Each test set is one of the three adversarial test sets (adversarial code generator applied on the original test set) on one of the 5 embedding-tasks.  

Table \ref{rq2_1_table} reports the raw scores as well as normalized improvements between the retrained models and the original models.  The first observation from this table is that all three retrained models can improve the robustness of the original models, by improving the performance of the downstream tasks in the presence of adversarial sample.

However, we can also observe that our GM method is significantly better, in improvements, compared to the other two alternatives. The GM method's median and mean normalized improvements  compared to the original model is 21.54\% (compared to 3.08\% and 4.84\%, for 1-Time and 5-Times, respectively) and 23.05\% (compared to 3.11\% and 6.39\%, for 1-Time and 5-Times, respectively). 

We also ran two Mann-Whitney U-tests to compare the distributions of reported normalized improvements for each paired comparison (1-Time vs. GM and 5-Times vs. GM). The results show that in both cases the p\_values are smaller than 0.05 ((0.0002 and 0.00142 respectively)) and thus the observed differences are statistically significant, as well.

\noindent\textbf{RQ2.2. Does re-training damage the original model performance, on the original test set?}

A further important question is whether augmenting the training set with adversarial test data will damage the model scores on the regular data (original test set) or not? In other words, we don't want the robustness process introduced in the work to only be useful on adversarial samples, but we rather need a model that not only is as good as the original model on the regular data but better than the original model on adversarials. RQ2.1 showed that our GM techniques properly takes care of the latter part. Thus we only need to check the former in RQ2.2.

Table \ref{rq2_2_table} reports all the scores and their normalized changes (see ``Improvement\%'') for the five embedding-tasks under study, when tested on their original test sets. Each ``Improvement\% is the normalized difference between the score of a retrained model (using of the three techniques for augmenting the training set) and the score of the original model (trained in the original training set).

We can observe that all retrained models negatively affect the original performance of the model on the original test sets. So none are ideal! However, the retrained models by GM adversarials have the least negative impact (median normalized decrease of performance is 3.56\% whereas 1-Time and 5-Times medians are at 7.31\% and 6.39\%, respectively). Therefore, we conclude that our approach reduces the performance of the original model on the regular data by a median of 3.56\% but since it improves the model robustness to adversarials by a median of 21.54\%, it can be a good strategy to implement especially if the robustness is critical for a particular task.

\begin{tcolorbox}[size=title,colback=white]
{\textbf{Answer to RQ2}: 
   Retraining embedding models using our approach improves the downstream tasks' performance on adversarial data sets by median of 21.54\%. This improvement is more than that of the alternatives. The retraining by our adversarial generation strategy also has the least negative impact on the performance on regular test-data (median of 3.56\%) compared to alternatives. 
}
\end{tcolorbox}

\subsection{Threats to Validity}
In terms of construct validity, we have used existing implementations of embedding models and reused their tools and dataset to avoid implementation biases. 
Regarding internal validity and to avoid confounding factors when drawing conclusions, we investigated three different embedding models and several downstream tasks to make sure the effects are not due to a particular model or task.  

With respect to conclusion validity, we repeat the GM algorithm 100 times with different random seed when selecting a refactoring operator to apply and report the median of the results, to avoid being the effect of randomness in the outputs. Note that the two baselines did not need this since they apply all refactoring operators in their pool.  In addition, while comparing our technique with two alternatives in RQ2.1 we ran two Mann-Whitney U-tests to make sure our conclusions are statistically significant. However, we could not do the same for RQ1 and RQ2.2 since we only had 5 observations per technique (RQ2.1 has 15 observations per technique). 

Finally, in terms of external validity, we have used three main embedding models and all their downstream tasks. However, more applications may also need to be tested through this experiment, in the future. I addition we only experimented with Java language code snippets as the input. It worth adding more programming languages like Python and C\# to see whether the performance of the refactoring-based adversarial examples is dependant on the programming language or not.

\section{Related Work} 
\label{relatedwork}
Rabin {\it et al.} \cite{rabin2020evaluation} and Ramakrishnan {\it et al.} \cite{ramakrishnan2020semantic} have used refactoring to generate adversarial samples for robustness testing of source-code based DNNs, specifically Code2vec (C2V)\cite{alon2019Code2vec} and Code2seq (C2S)\cite{alon2018Code2seq}.

Rabin {\it et al.} \cite{rabin2020evaluation} apply semantics preserving program transformations to produce new programs using refactoring methods on which they expect models to keep their original predictions and report the prediction change rate. However, they have not retrained the model using adversarial test input to see if any improvement happens. 

Ramakrishnan {\it et al.}\cite{ramakrishnan2020semantic} focused on the number of changes (refactorings) applied to each test input. As stated, they have tried different values of K, which defines the number of times refactoring operators are going to be applied on the given. They reported that K = 5 is the best based on their experiment on the Code2seq model. 

Bielik {\it et al.} \cite{bielik2020adversarial} also focused on creating accurate and robust models for codes. They found two main challenges in determining the robustness of the models trained for code, 1) the programs are significantly large and structured compared to images or natural language, and 2) computing the correct label for all samples is challenging. To address these two challenges, they first find the program's parts relevant for the prediction and then let the model make predictions for some of the samples and not force the model to make predictions for all the samples. They also train different models instead of just one so that each model is simpler and thus easier to train robustly.

In Want {\it et al.} work \cite{wang2019coset}, they introduce a benchmark framework called \textit{COSET} for evaluating neural embedding programs proposed for the software classification task. They also show that COSET program transformations can identify the programming language characteristics, the program code, or the program execution that causes the accuracy drop.

Yefet {\it et al.} \cite{yefet2019adversarial} presented a general white-box technique called \textit{DAMP}, which targeted attacks of models using adversarial examples for source code. DAMP works by the prediction concerning the model’s inputs. While the model weights are holding constant, the gradients slightly modify the input code. They finally show that their attack is effective across three neural architectures.

Zhang {\it et al.} \cite{zhang2020generating} introduce a model for generating adversarial samples for source code called \textit{MHM}. MHM generates adversarial examples by iteratively restricting the perturbations on the examples to satisfy programming languages' constraints. The results show that the generated adversarial examples attack the subject models with high attack and validity rates. They also show that with adversarial training, the classification model performance improves.

In our study, we have two contributions compared to related work: (a) we propose a new search-based testing framework to create code adversarial examples based on refactoring operators and mutation testing guidance, and (b) we retrain the models with the adversarial examples to improve the robustness of the models on multiple downstream tasks.

\section{Conclusion}
Providing robust, safe, and secure deep neural networks is one of the main challenges of current machine learning systems. In this work, we proposed a novel search-based testing technique for code embedding models to evaluate their robustness. The technique uses an iterative guided refactoring process to generate adversarial code snippets that can fool a downstream task, which uses the embedded code snippets. 
By retraining embedding models on augmented training sets (enriched by the adversarial code snippets), we can improve their performance on the downstream tasks such as code captioning, method name prediction, and code search, which resulted in up to 17\% improvements compared with the state-of-the-art methods.

\section*{Acknowledgments}
This work was supported in part by JSPS KAKENHI Grant No.20H04168, 19K24348, 19H04086, and JST-Mirai Program Grant No.JPMJMI18BB, Japan

\balance

\bibliography{ref}
\bibliographystyle{ieeetr}

\end{document}